\documentclass[aip,apl,letterpaper,amsmath,amssymb,reprint]{revtex4-1}
\usepackage{graphicx}
\usepackage{microtype}

\begin{document}
\title{Temperature dependence of the spin Hall angle and switching current in the nc-W(O)/CoFeB/MgO system with perpendicular magnetic anisotropy}
\author{L. Neumann, D. Meier, J. Schmalhorst, K. Rott, G. Reiss, M. Meinert}
\email{meinert@physik.uni-bielefeld.de}
\affiliation{Center for Spinelectronic Materials and Devices, Department of Physics, Bielefeld University, D-33501 Bielefeld, Germany}

\date{\today}

\begin{abstract}
We investigated the temperature dependence of the switching current for a perpendicularly magnetized CoFeB film deposited on a nanocrystalline tungsten film with large oxygen content: nc-W(O). The spin Hall angle $|\Theta_\mathrm{SH}| \approx 0.22$ is independent of temperature, whereas the switching current increases strongly at low temperature. We show that the nc-W(O) is insensitive to annealing. It thus can be a good choice for the integration of spin Hall driven writing of information in magnetic memory or logic devices that require a high-temperature annealing process during fabrication.
\end{abstract}

\maketitle

\begin{figure}[b]
\includegraphics[width=8.6cm]{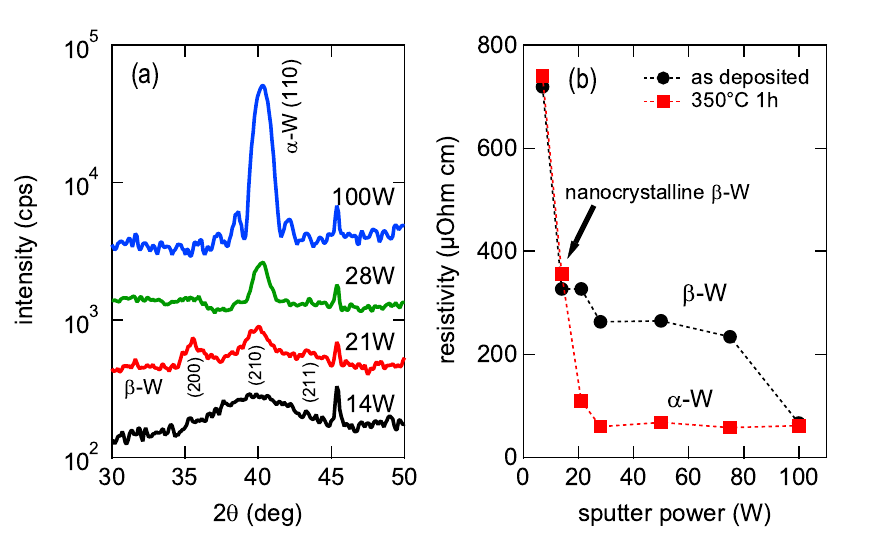}
\caption{\label{fig:xrd_vs_power} X-ray diffraction patterns (a) and electrical resistivities (b) for 10\,nm thick W films sputtered with different powers in an Ar + 0.5\% O$_2$ atmosphere.}
\end{figure}

The ability of spin-orbit torques, and in particular of the spin Hall effect\cite{Dyakonov1971, Hirsch1999, Hoffmann2013} (SHE), to generate pure spin currents that can be used to manipulate the magnetization of an ultrathin magnetic  film has triggered very active research of the spintronics community on this novel field. The spin current generated through the SHE in a heavy metal film exerts a spin transfer torque (STT) on an adjacent magnetic film which can be strong enough to excite ferromagnetic resonance dynamics and even magnetization reversal.\cite{Liu2011} The pivotal quantity of the SHE is the spin Hall angle $\Theta_\mathrm{SH} = j_\mathrm{s} / j_\mathrm{c}$ describing the ratio of the spin current $j_\mathrm{s}$ and the orthogonal charge current $j_\mathrm{c}$. After initial successful demonstrations of current-induced magnetization switching in perpendicularly magnetized Ta / CoFeB ($\Theta_\mathrm{SH} = -0.12$) and Pt / Co ($\Theta_\mathrm{SH} = 0.07$) film stacks,\cite{Liu2012a, Liu2012b, Avci2012} a search for materials with larger spin Hall angles started. Being a spin-orbit interaction derived effect, it is expected to be largest in heavy metals or lighter metals with heavy metal impurities.\cite{Tanaka2008, Niimi2012} The highest values for a metal were observed for $\beta$-W with the A15 structure; values between $|\Theta_\mathrm{SH}| = 0.33$ and $|\Theta_\mathrm{SH}| = 0.4$ were found after careful optimization of the film growth conditions to retain the metastable A15 structure.\cite{Pai2012, Hao2015a} Deliberate oxygen incorporation proved successful to stabilize $\beta$-W(O) films with giant spin Hall angle of $\Theta_\mathrm{SH} = -0.5$.\cite{Demasius2016, Weerasekera1994} At the same time, high oxygen content above 25\% was shown to induce a nanocrystalline growth of the W film.\cite{Demasius2016} However, the $\beta$-W phase is sensitive to high temperatures and may thus be unsuitable for integration into device fabrication processes that require sustained annealing at high temperature, e.\,g., during CMOS manufacturing.\cite{OKeefe1996} In the present article we focus on nanocrystalline W which is stable against high-temperature annealing and investigate its utility for current-induced magnetization switching.

The films were prepared by dc magnetron sputtering in a 2-inch confocal co-sputtering system with a target-to-substrate distance of 10\,cm and a source inclination of 30$^\circ$. The base pressure of the system was below $5 \times 10^{-9}$ mbar.  For the controlled and reproducible deposition of $\beta$-W(O) an Ar flow of 40\,sccm with 0.2\,sccm O$_2$ (0.5\%) added was used and the working pressure was set to $3 \times 10^{-3}$\,mbar. Because the amount of oxygen incorporated into the film depends on the deposition rate of the W, the sputtering power was varied between 7\,W and 100\,W (deposition rate: $3.2\,\mathrm{pm}/\mathrm{Ws}$ without O$_2$ flow) in order to identify the optimal deposition conditions for nanocrystalline $\beta$-W(O) by x-ray diffraction and resistivity measurements. In all cases, natively oxidized GaAs substrates were used. As shown in Figure \ref{fig:xrd_vs_power}, the W deposited at 100\,W is in the $\alpha$ phase with very smooth interfaces as indicated by the pronounced Laue oscillations. At the same time, the resistivity is low (around 60\,$\mu \Omega \mathrm{cm}$) and does not change upon annealing at 350$^\circ$C. Powers between 28\,W and 75\,W lead to mixed $\alpha$ and $\beta$ phase films which completely transform to the $\alpha$ phase upon annealing. The film deposited at 21\,W is close to single phase $\beta$-W(O). However, still a transformation into $\alpha$-W is observed after annealing. At 14\,W, nanocrystalline growth with a grain size of about 1.8\,nm is observed. This film as well as a highly oxidized film deposited at 7\,W show only marginal changes in resistivity after annealing. In the following we focus on the nanocrystalline W (nc-W(O)) deposited at 14\,W. Film stacks of nc-W(O) 6 / CoFeB 1.3 / MgO 1.6 / Ta 1 (all thicknesses in nm) were and annealed at 300$^\circ$C for 1h in a 0.7\,T perpendicular magnetic field to maximize the perpendicular magnetic anisotropy. After annealing, a sample was patterned into $4.2 \times 22\,\mu\mathrm{m}^2$ Hall bar structures with contact lines by electron beam lithography and ion beam milling. The oxygen content of this film was estimated by a density measurement with x-ray reflectivity. We found about 12\,g/cm$^3$, which corresponds to 40\% oxygen content, assuming the A15 structure of $\beta$-W where oxygen atoms substitute for W atoms. Additional x-ray diffraction and reflectivity measurements with thicker W(O) films confirm that the amount of oxygen incorporated into the film decreases with increasing film thickness, as reported by Demasius \textit{et al}.\cite{Demasius2016}

\begin{figure}[t]
\includegraphics[width=8.6cm]{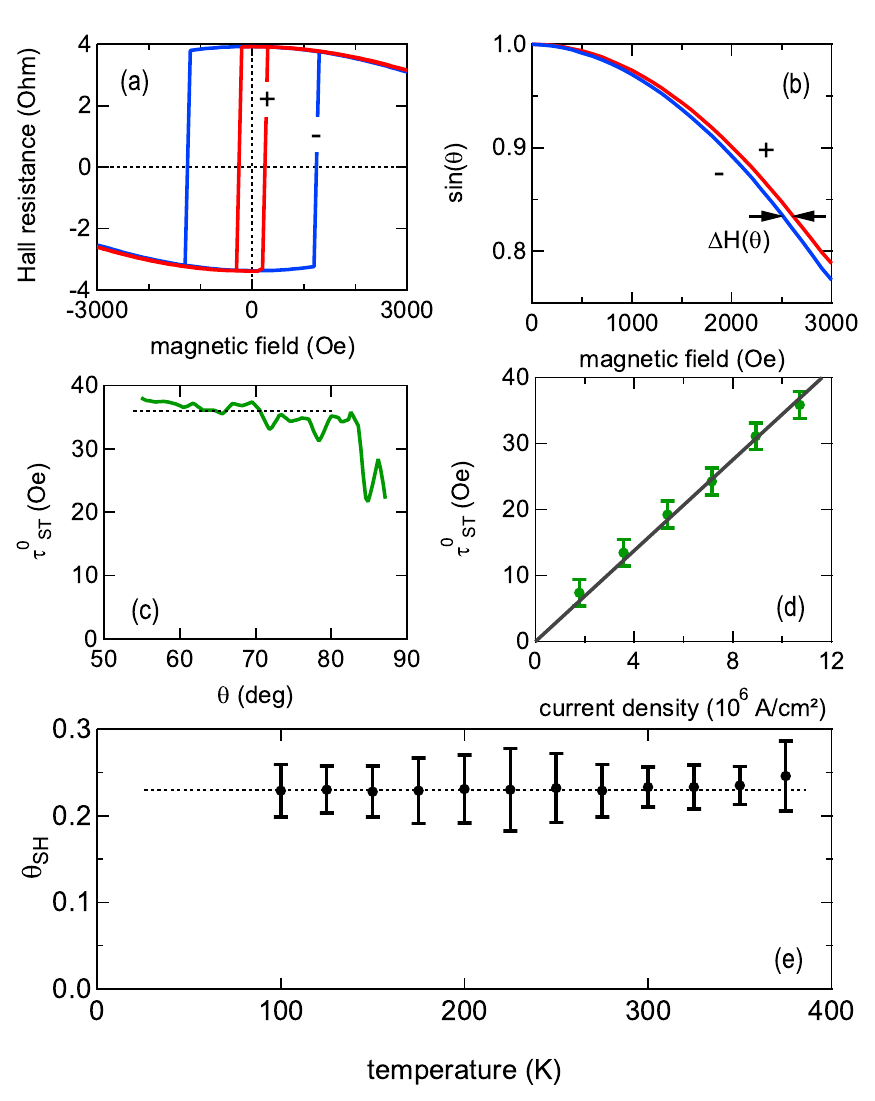}
\caption{\label{fig:SHE_analysis}(a): In-plane magnetic field loops monitoring the Hall resistance $R_\mathrm{H}$ ($\beta = 13^\circ$) for currents of $\pm3$\,mA at 275\,K. (b): $\sin(\theta) = R_\mathrm{H}(H_x) / R_\mathrm{H}(0)$ extracted by removing the offset in $R_\mathrm{H}$ and the switching from data in (a). Additionally, values for positive and negative fields were averaged. (c): $\tau_\mathrm{ST}^0 (\theta)$ extracted from $\Delta H$ in (b) using equation \ref{eq:torque}. $\tau_\mathrm{ST}^0$ is obtained by averaging $\tau_\mathrm{ST}^0(\theta)$ for $\theta < 80^\circ$ (dotted line). (d): $\tau_\mathrm{ST}^0$ obtained for different current densities $j_c$. $\tau_\mathrm{ST}^0 / j_c$ is obtained by a linear fit. (e): Spin Hall angle evaluated using equation \ref{eq:SHE} as a function of temperature.}
\end{figure}

\begin{figure}[t]
\includegraphics[width=8.6cm]{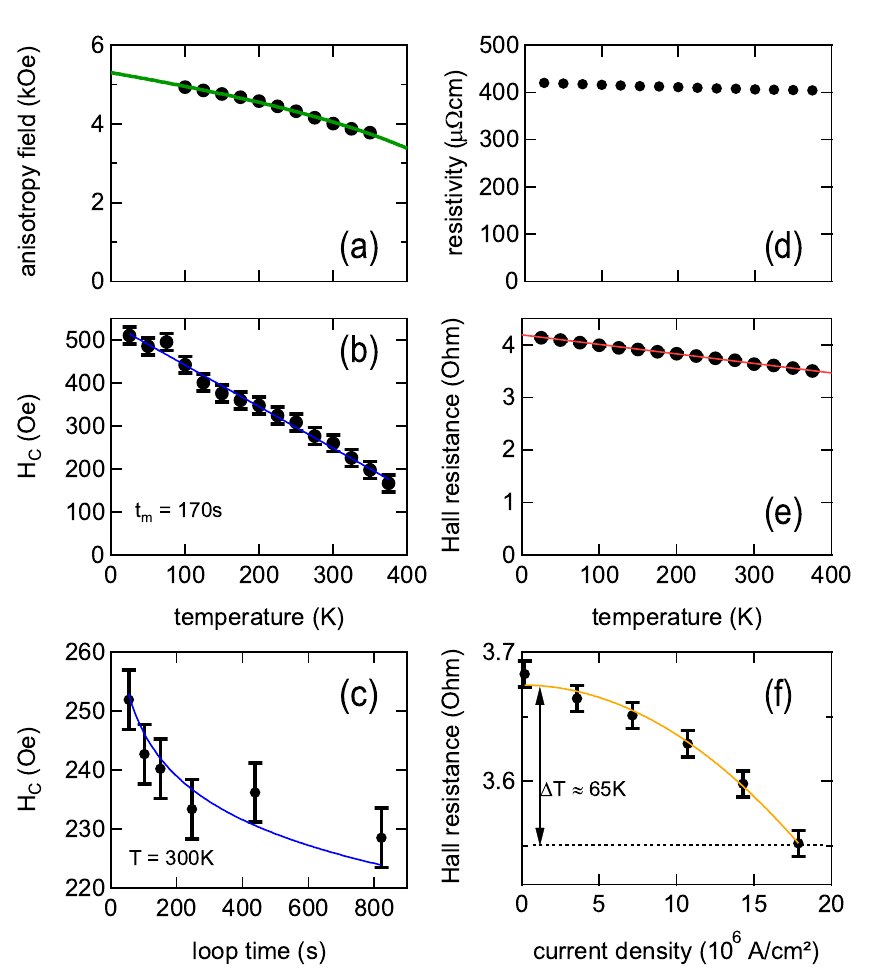}
\caption{\label{fig:magnetics_vs_T}(a): Anisotropy field as a function of temperature with a fit to the equation $H_\mathrm{an}^0(T) = H_0 [1-T/T_\mathrm{B}]^\beta$ with $\beta = 0.33$, and $T_\mathrm{B} = 536$\,K. (b) and (c): Coercive field as a function of temperature and fit to equation \ref{eq:depinning}; Coercive field as a function of loop measurement time and fit to equation \ref{eq:depinning}. (d):  Electrical resistivity as a function of temperature. (e) and (f):  Hall resistance as a function of temperature with line fit; Hall resistance as a function of Hall bar current density at 300\,K with parabolic fit.}
\end{figure}

Current-induced switching and spin Hall torque measurements were performed in a closed-cycle He cryostat between 100\,K and 375\,K with a magnetic field up to $\pm 3000$\,Oe. The spin Hall angle was measured with the torque method of Reference \onlinecite{Liu2012a} where the $z$-component of the magnetization was monitored with the anomalous Hall effect. The magnetic field was applied under an angle of $\beta =13^\circ$ with respect to the sample surface to obtain an out-of-plane component that suppresses domain formation and keeps the magnetization in the coherent rotation regime for sufficiently small current densities. The anomalous Hall effect was measured as a function of temperature and positive/negative current densities to obtain $R_H(H_x) / R_H(0) = \sin(\theta)$ as a measure for the magnetization angle with respect to the sample surface $\theta$. The total torque acting on the magnetization can be written as $\vec{\tau}_\mathrm{tot} = \vec{\tau}_\mathrm{ST} + \vec{\tau}_\mathrm{ext} + \vec{\tau}_\mathrm{an}$, where $\vec{\tau}_\mathrm{ST} = \tau_\mathrm{ST}^0 (\hat{m} \times (\hat{\sigma} \times \hat{m}))$ is the anti-damping (Slonczewski-like) spin-transfer torque per unit moment due to the spin Hall effect, $\vec{\tau}_\mathrm{ext}$ is the torque due to the external magnetic field, and $\vec{\tau}_\mathrm{an}$ is the torque due to the anisotropy field. $\hat{\sigma}$ denotes the direction of the spin polarization of the spin current $j_s$. With the stability condition for the magnetization $\vec{\tau}_\mathrm{tot} = 0$ this equation can be rewritten in a scalar form as\cite{Liu2012a}
\begin{align}\label{eq:torque}
H_+(\theta) - H_-(\theta) &= \Delta H(\theta) = -2\tau_\mathrm{ST}^0 / \sin(\theta-\beta),\nonumber\\
H_+(\theta) + H_-(\theta) &= -2H_\mathrm{an}^0 \sin(\theta)\cos(\theta) / \sin(\theta-\beta),
\end{align}
with the external field values $H_+(\theta)$ and $H_-(\theta)$ to obtain the same angle $\theta$ of the magnetization with positive or negative current, respectively. Individual steps of the analysis are described in Figure \ref{fig:SHE_analysis} (a) - (d), which demonstrates the expected proportionality between $\tau_\mathrm{ST}^0$ and the charge current density. The main result of the analysis is the spin Hall angle 
\begin{equation}\label{eq:SHE}
\Theta_\mathrm{SH} = \frac{2e}{\hbar} \frac{\tau_\mathrm{ST}^0}{j_c} M_\mathrm{s} t_\mathrm{F}
\end{equation}
that is essentially independent of temperature with $\Theta_\mathrm{SH} = 0.22 \pm 0.03$ as shown in Figure \ref{fig:SHE_analysis} (e). The torque per unit current density flowing through the nc-W(O) layer is $\tau_\mathrm{ST}^0 / j_c = 5.64 \times 10^{-6}\,\mathrm{Oe}\,\mathrm{cm}^2/\mathrm{A}$. The electrical resistivities of the full stack and CoFeB were taken into account to compute the actual current density through the nc-W layer, where we assume $\rho_\mathrm{CoFeB}\approx 150\,\mathrm{\mu\Omega cm}$. The resulting resistivity of the nc-W(O) with 6\,nm thickness is about $600\,\mu\Omega\mathrm{cm}$, which is significantly higher than the value obtained at 10\,nm thickness. This is in line with the previous finding of an increased oxygen content at lower film thickness, hence giving rise to an enhanced resistivity. The magnetization of the CoFeB layer was measured with an alternating gradient magnetometer at room temperature [$M_\mathrm{s} = (1000 \pm 100)\,\mathrm{kA/m}$] and was assumed to be independent of temperature in this analysis. The value of $\Theta_\mathrm{SH}$ of the nc-W(O) is considerably reduced in comparison with crystalline $\beta$-W, but it is still larger than the highest values obtained for $\beta$-Ta. The reduction of the spin Hall angle in nc-W(O) was already implied in the results discussed by Demasius \textit{et al.}  and our value is consistent with their value from FMR line-shape analysis.\cite{Demasius2016} The fact that $\Theta_\mathrm{SH}$ is independent of temperature within the measurement accuracy is not surprising in view of the electrical resistivity of the film stack, which is also nearly constant, cf. Figure \ref{fig:magnetics_vs_T}\,(d). Recent studies on the temperature dependence of $\Theta_\mathrm{SH}$ in Ta / CoFeB or YIG / Pt  systems found that $\Theta_\mathrm{SH} \propto \rho_\mathrm{NM}$ or $\Theta_\mathrm{SH} \propto \rho_\mathrm{NM}^2$,\cite{Hao2015b, Wang2016} very similar to the clean and dirty limits of the anomalous Hall effect.\cite{Nagaosa2010}

\begin{figure}[t]
\includegraphics[width=8.6cm]{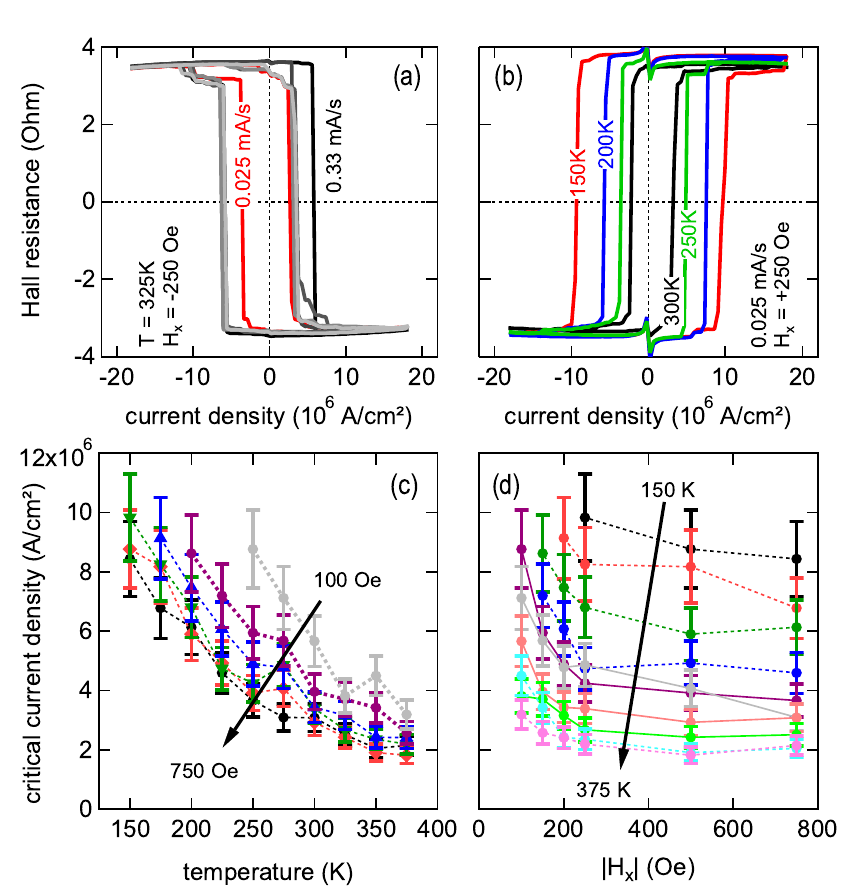}
\caption{\label{fig:critical_current}Analysis of current-induced magnetization reversal loops. (a): Repeated current loops with fast ramp-rate (black to gray) and with slow ramp-rate (red). (b): Current loops at different temperatures. Note the reversed sign of the switching compared to (a) due to the  reversed in-plane field. (c): Critical current densities for magnetization reversal as a function of temperature, taken at different $|H_x|$ with a ramp rate of 0.025\,mA/s. (d): Sama data as (c) as a function of  $|H_x|$.}
\end{figure}

In Figure \ref{fig:critical_current} we show the results of the experiments on current-induced magnetization reversal with an external in-plane assist-field $H_x$. Switching was observed to be a stochastic process that depends on the current sweep-rate as shown in Figure \ref{fig:critical_current}(a): with larger sweep-rate, the switching current density is larger and the two switching current densities for up/down or down/up switching can be severely different. Also, switching through many intermediate states and incomplete reversal is more often observed at higher sweep-rate, in particular for small $H_x$. As shown in Figure \ref{fig:critical_current}, the critical current density increases significantly at low temperature. At temperatures below 150\,K, no reliable switching was observed with the maximum current density of $18 \times 10^6\,\mathrm{A/cm^2}$ that we have applied. Higher current densities were not applied to avoid excessive heating of the device. With the observed switching current densities below $10 \times 10^6\,\mathrm{A/cm^2}$ the heating of the current channel was estimated to be below 30\,K (see detailed analysis below). The critical current density after averaging positive/negative currents and positive/negative $H_x$ is shown in Figures \ref{fig:critical_current} (c) and (d) as functions of temperature and in-plane assist-field $H_x$. For all values of $H_x$ between 100\,Oe and 750\,Oe the critical current density increases as temperature is decreased. At elevated temperatures below 375\,K, the slope of the critical current density as a function of temperature levels off. The switching phase diagram obtained from the data in Figure \ref{fig:critical_current} (c) is given in Figure \ref{fig:critical_current}(d): a significant increase in the critical current density for $H_x < 200$\,Oe is observed for all temperatures.

The variation of the switching current density can not be explained in terms of a variation of the spin Hall angle in the nc-W(O) / CoFeB / MgO system. To gain further insight into the switching behavior and the origin of the temperature dependence, we plot in Figure \ref{fig:magnetics_vs_T} the temperature dependencies of the anisotropy field $H_\mathrm{an}^0$, the coercive field $H_c$, the electrical resistivity $\rho$, and the anomalous Hall resistance $R_\mathrm{H}$. Remarkably, the anisotropy field depends only weakly on temperature for low temperatures and has a temperature dependence that resembles a typical $M(T)$ curve (see caption for fit parameters). Using the macrospin expression for the critical current density in a perpendicularly magnetized magnetic thin film, $j_\mathrm{c}^\mathrm{SH} = \frac{2e}{\hbar}\frac{M_\mathrm{s} t_\mathrm{F}}{\Theta_\mathrm{SH}}(H_\mathrm{an}^0/2 - H_x/\sqrt{2})$,\cite{Lee2013} yields a current density two orders of magnitude larger than observed. Clearly, the magnetization decays into domains during the magnetization reversal as $H_\mathrm{c} \ll H_\mathrm{an}^0$ for all temperatures (Figure \ref{fig:magnetics_vs_T}\,(a) and (b)). The coercive field measured with the magnetic field perpendicular to the sample surface shows a linear variation with temperature (Figure \ref{fig:magnetics_vs_T}\,(b)), which is typical of thermally activated domain wall depinning.\cite{Lee2014} Magnetization loops with different field sweep rates were taken to determine the measurement time dependence of the coercive field, Figure \ref{fig:magnetics_vs_T} (c). Both curves can be fitted with a thermal activation model,\cite{Lee2014, Kvhalkovskiy2013}
\begin{equation}\label{eq:depinning}
\left< H_p\right> = H_{p,0}^z \left( 1 - \left[ \frac{k_B T}{E_p} \ln\left(\frac{f_0 t_m}{\ln 2}\right) \right] \right)
\end{equation}
for the depinning field $H_p$. Here, $H_{p,0}^z$ describes the depinning field in the absence of thermal fluctuations, $E_p$ is the pinning energy, $f_0$ is the attempt frequency, assumed to be 1\,GHz,\cite{Kvhalkovskiy2013} and $t_m$ is the measurement time (taken as the total time for a field loop of $\pm 300$\,Oe). From the temperature dependence shown in Figure \ref{fig:magnetics_vs_T}\,(b) one obtains $E_p = (1.26 \pm 0.03)$\,eV and $H_{p,0}^z = (539 \pm 7)$\,Oe. The same parameters also describe the measurement time dependence of the coercive field shown in in Figure \ref{fig:magnetics_vs_T}\,(c) very well. This confirms the thermally activated domain wall depinning as the mechanism responsible for the magnetization reversal in an out-of-plane field loop experiment. The thermal stability factor $\Delta = E_p / k_\mathrm{B}T \approx 50$ is large enough to guarantee stable data retention for 10 years in a memory device based on the present thin film system. However, the corresponding expected critical current densities are of the order $j_\mathrm{c}^\mathrm{SH} = \frac{2e}{\hbar}\frac{M_\mathrm{s} t_\mathrm{F}}{\Theta_\mathrm{SH}}H_{p} \approx 5 \times 10^7\,\mathrm{A/cm^2}$, still an order of magnitude larger than observed. This indicates that details of the domain nucleation and expansion in the presence of the current and the in-plane field differ from the field-driven magnetization reversal and are important for the quantitative understanding of the critical current density.

The influence of heating on the $j_\mathrm{c}(T,H_x)$ curves was studied with the anomalous Hall resistance amplitude as a function of temperature for low current density and as a function of current at room temperature. The variation of $R_\mathrm{H}$ with temperature is essentially linear with a small negative slope of $-0.0018\,\Omega/\mathrm{K}$. From the parabolic variation of the Hall resistivity with current density we estimate $\Delta T < 65\,\mathrm{K}$ for the highest current densities of $16.9 \times 10^6\,\mathrm{A/cm}^2$ used during the current loops. At the highest switching current densities (around $10 \times 10^6\,\mathrm{A/cm}^2$) the temperature increase is $\Delta T < 30\,\mathrm{K}$, just above the temperature step used for the measurements. Thus heating does contribute to a reduction of the switching current, however only to a minor degree. 

Finally, we comment on the switching observed in Fig. \ref{fig:SHE_analysis}. The magnetic field switching observed here comes from the out-of-plane component of the canted magnetic field with $\beta = 13^\circ$.  However, the two curves for $+3$\,mA and $-3$\,mA show very different coercive fields. The associated current density of $9.6 \times 10^6\,\mathrm{A/cm}^2$ is well above the critical current density for the magnetization switching. This modification of the coercive field has the symmetry of the STT leading to the current-induced switching (cf. Figures \ref{fig:critical_current}\,(a) and (b)): The positive (negative) current stabilizes the positive (negative) magnetization direction at positive in-plane magnetic field as well as the negative (positive) magnetization direction at negative in-plane magnetic field, giving rise to lowered (enhanced) coercive field. 

To summarize, we observed that the spin Hall angle is independent of temperature in the nc-W(O) / CoFeB / MgO system. Nevertheless, the critical current density for the current-induced magnetization switching increases greatly with temperature, which is related to the increased coercive field, but can not quantitatively explained by this. The domain nucleation, depinning, and expansion in the presence of the in-plane current need to be understood in more detail to quantitatively explain the temperature dependence of the switching current density. The system presented here has significant potential for applications, thanks to its superior temperature stability as compared to crystalline $\beta$-W films with lower oxygen content, and a still high spin Hall angle.

\end{document}